\documentclass[11pt]{article}              

\usepackage[margin=25mm]{geometry}
\usepackage{graphicx}
\usepackage{colortbl}
\usepackage{subfigure}
 
\usepackage{fancyhdr}
\usepackage{amssymb,amsfonts,amsmath,amsthm}
\usepackage{natbib}

\usepackage{authblk}
	
\graphicspath{{./},{./Figures/}}

\usepackage{notations}
\usepackage{multirow}
\usepackage{hyperref}
\usepackage{booktabs}
\begin{document}
\title{Reliability-based design optimization of shells with uncertain
	geometry using adaptive Kriging metamodels} 

\author[1]{V. Dubourg}  \author[2,3]{J.-M. Bourinet} \author[4]{B. Sudret}

\affil[1]{Phimeca Engineering, Cournon d'Auvergne, France}
\affil[2]{Universit\'e Clermont Auvergne, Sigma Clermont, Institut Pascal, Clermont Ferrand, France}
\affil[3]{CNRS, UMR 6602, Institut Pascal, Aubi\`ere, France}
\affil[4]{Institute of Structural Engineering, ETH Z\"urich, Z\"urich, Switzerland}

\date{}
\maketitle

\abstract{Optimal design under uncertainty has gained much attention in the past ten years due to the ever increasing need for manufacturers to build robust systems at the lowest cost.Reliability-based design optimization (RBDO) allows the analyst to minimize some cost function while ensuring some minimal performances cast as admissible failure probabilities for a set of performance functions. In order to address real-world engineering problems in which the performance is assessed through computational models (\eg finite element models in structural mechanics) metamodeling techniques have been developed in the past decade.
	This paper introduces adaptive Kriging surrogate models to solve the
	RBDO problem. The latter is cast in an augmented space that ``sums up'' the range of the design space and the aleatory uncertainty in the design parameters and the environmental conditions.
	The surrogate model is used \emph{(i)} for evaluating robust estimates of the failure probabilities (and for enhancing the computational experimental design by adaptive sampling) in order to achieve the requested accuracy and \emph{(ii)} for applying a gradient-based optimization algorithm to get optimal values of the design parameters.
	The approach is applied to the optimal design of ring-stiffened cylindrical shells used in submarine engineering under uncertain geometric imperfections.
	For this application the performance of the structure is related to buckling which is addressed here by means of a finite element solution based on the asymptotic numerical method. \\[1em] 

}

\maketitle

%%%%%%%%%%%%%%%%%%%%%%%%%%%%%%%%%%%%%%%%%%%%%%%%%%%%%%%%%%%%%%%%%%%%%%%%%%%%
\section{Introduction}

Shell structures occupy a predominant part of our landscape \cite[see \textit{e.g.}][for a review of their applications]{Ramm2004}.
They owe this predominance to their curvature which allow them to withstand large transverse loading by a membrane-dominated stress state.
As a result, they can be used for building large-span shelters such as roofs, fuselages or boat and submarine hulls without requiring too many intermediate supports such as stiffening beams or rims.
Nonetheless, as many optimized and therefore slender structures, the strength of thin shells also exhibits a significant sensitivity with respect to geometrical, material and other environmental conditions which are typically unknown to some extent.
\par

Early work on the elastic stability of slender structures (such as beams, plates or shells) is often attributed to Euler in 1744, although most of the theoretical concepts in practice today for shells are due to Lorentz, Timoshenko and Southwell in early nineties.
In parallel to theoretical advances, experimental studies revealed embarrassing discrepancies between the predicted buckling loads and those obtained from real tests.
\citet{Koiter1945} was certainly the first researcher to point out that these discrepancies are mostly explained by the imperfect geometry, boundary conditions and material properties of the experimental specimens.
This premise is now fully acknowledged by the whole community of engineers and scientists in structural mechanics in the light of other studies by \citet{Arbocz1969,Singer1971,Singer1995} amongst others.
The reader may refer to \citep{Bazant2000} for a review of works in the field of stability of structures with an emphasis on anelastic structures and to the recent paper from \citet{Elishakoff2012} for a detailed history of works on elastic stability of shells.
\par

A key aspect of these imperfections though is that they are extremely varying in terms of shape and amplitude.
Hence, for the sake of structural safety, designers have to account for extreme and fortunately unlikely imperfections.
A common practice is to assume a given shape in the calculations corresponding to the worst case structural strength and then resort to advanced numerical schemes in order to justify the design.
However, this approach, referred to as the worst case approach in the sequel, introduces an unknown degree of conservatism which may not suit the safety requirements imposed by stakeholders.
\par

As early suggested by \citet{Bolotin1958} in his pioneering works, it is argued that a better solution may be obtained by means of statistical methods and that the design of imperfect shells necessarily falls under a probabilistic formulation.
Several imperfection surveys were later carried out in order to assess the statistical properties of imperfections present in both small and large-scale shells.
These statistics such as those gathered in the imperfection data bank \citep{Arbocz1982} were then introduced into stochastic buckling analysis by researchers.
\citet{Elishakoff1979} was the first researcher to use random initial imperfections of compressed cylindrical shells in a Monte Carlo analysis.
The initial imperfections were expanded in double Fourier series and the Fourier coefficients were considered as Gaussian random variables.
Such a representation of the random imperfections is also used in further studies carried out by the same authors, see \eg \citep{Elishakoff1982,Elishakoff1987,Arbocz1995,Arbocz2005}.
This double Fourier representation was later investigated by other researchers with an effort to reduce the number of random Fourier coefficients in order to alleviate the cost of the stochastic analysis \citep{Noirfalise2009PhD,Kriegesmann2010,Kriegesmann2011}. 
In the works of Schenk and Schu{\"e}ller \citep{Schenk2003,Schenk2007}, the geometric imperfection was modeled as a two-dimensional univariate non-homogeneous Gaussian random field by means of a Karhunen Lo{\`e}ve expansion.
This representation is also used in references \citep{Craig2008,Noirfalise2009PhD,Dubourg2009ICOSSAR}.
As an alternative technique, the spectral representation method was applied in several works for modeling two-dimensional univariate random fields, see \eg the work by \citet{Stefanou2004} which models homogenous Gaussian random fields representing spatially-varying material properties (Young's modulus and Poisson's ratio) and thickness imperfections.
The spectral representation method is also used in conjunction with an autoregressive moving average technique with evolutionary power spectra in \citep{Papadopoulos2005} for modeling non-homogenous Gaussian random fields representing spatially-varying Young's modulus in addition to geometric and thickness imperfections.
A similar approach based on non-Gaussian translation fields is adopted by \citet{Papadopoulos2005}, which puts the emphasis on the influence of the Gaussianity/non-Gaussianity assumption on the results.
A recourse to evolutionary power spectra estimated by means of a moving window averaging technique is also found in \citep{Broggi2011} for non-homogenous random fields representing geometric and thickness imperfections.
Most of the works reported in the literature focused on metallic shells with geometric imperfections possibly combined with spatially-varying material properties and thickness imperfections.
A few research studies were also conducted on anisotropic composite shells \citep{Chryssanthopoulos1995,Kriegesmann2010,Broggi2011,Kriegesmann2011} characterized by larger imperfections due to their complex manufacturing processes.
For a more realistic treatment of imperfections, some other sources of random imperfections were additionally incorporated in probabilistic buckling studies such as those arising from a non-uniform distribution of the axial loading \citep{Papadopoulos2007} or those coming from the application of uncertain boundary conditions \citep{Schenk2007}.
It is of importance to mention that most of the models used for describing the geometric imperfections in the above cited references were identified from the experimental data available in the imperfection data bank.
For thickness imperfections and spatially-varying material properties, the parameters of the random fields are assumed to have specified values and they are sometimes varied in a parametric analysis.
\par

In early probabilistic studies, buckling loads were computed by means of Koiter's theory \citep{Elishakoff1979,Elishakoff1982}.
More refined numerical solutions based on a multimode analysis were later used by \citet{Elishakoff1987} and \citet{Arbocz2005}.
A recourse to a nonlinear finite element (FE) model is often advocated for an enhanced accuracy on the limit loads.
FE-based probabilistic approaches were carried out with STAGS in \citep{Arbocz1995,Schenk2003,Schenk2007}, ABAQUS in \citep{Kriegesmann2010,Broggi2011,Kriegesmann2011} and LS-DYNA in \citep{Craig2007,Craig2008} (note that this last code is specifically used in the context of dynamic buckling).
In other studies, some other researchers implemented specific shell elements for the purpose of their probabilistic buckling analysis.
The TRIC triangular shell element as described in \citep{Argyris2002} is used in all the works carried out at the National Technical University of Athens \citep{Stefanou2004,Papadopoulos2005,Lagaros2006,Papadopoulos2007,Papadopoulos2009}.
In their works, \citet{Noirfalise2009PhD} and \citet{Dubourg2011PhD} made a recourse to the B{\"u}chter and Ramm 8-node shell element \citep{Buchter1994} for their reliability analysis and reliability-based design optimization.
Most of the probabilistic buckling analyzes found in the literature assume linear elasticity of metallic shells, which is perfectly appropriate for the studied thin shells taken from the imperfection data bank such as the so-called A-shells.
A nonlinear behavior of the shell material is addressed in \citep{Papadopoulos2005,Lagaros2006,Dubourg2009ICOSSAR}.
The Young's modulus is considered as a random field in all these references.
The yield strength is taken as an additional random field independent of the Young's modulus one by Dubourg \etal in \citep{Dubourg2009ICOSSAR}.
\par

In several studies, the probabilistic analysis consisted in constructing limit load histograms with comparison to those obtained experimentally and analyzing the second-order statistics of these loads.
Note that the numbers of samples used in the FE-based Monte Carlo simulations of the reported references are most often in the order of a few hundreds.
Another direction followed by researchers has consisted in performing a structural reliability analysis of shells by imposing that their limit loads should be greater than a prescribed service load.
The earliest occurrences of such studies were based on the first-order second-moment method \citep{Elishakoff1987,Arbocz1995,Arbocz2005}.
The first-order reliability method (FORM) was later used in many works based on analytical models not listed here for the sake of brevity or on FE models, see \eg \citep{Bourinet2000IASSIACM,Dubourg2009ICOSSAR}.
For the specific case of imperfections modeled by random fields, a recourse to subset simulation method \citep{Au2001} is considered as the most suitable solution as investigated by \citet{Noirfalise2009PhD} and \citet{Dubourg2009ICOSSAR}.
For the purpose of improved designs, the optimization of shell subject to buckling based on FE models has also been of interest in recent years.
In \citep{Lagaros2006}, the weight of shells with random geometric imperfections and space-variant Young's modulus and thickness is minimized in the framework of reliability-based design optimization.
This work is based on a $(5+5)$-evolution strategy optimization algorithm and the probabilistic constraint is assessed by means of a crude MC with $1,000$ samples.
In \citep{Craig2008}, shells with stochastic imperfections are optimized in a dynamic buckling context.
Two optimization studies are performed: the first one aims at minimizing the weight of the shell with constraints on the average peak force and average internal energy, the second one aims at increasing the robustness \wrt the variations of the normal peak force with the same constraints.
The strategy used by \citeauthor{Craig2008} is to construct a quadratic polynomial response surface with a 96-sample MC simulation at each point of the design of experiments.
\par

%Mesh convergence study for a FE solution of a sufficient accuracy \citep{Schenk2003,Papadopoulos2005}

% sample # in MC
%Craig 100
%Schenk2003 250
%Schenk2007 50
%Kriegsmann2010 3000 
%Broggi2011 100
%Argyris2002 500
%Stefanou2004 500
%Papadopoulos2005 100
%Lagaros 2006 100

%The initial imperfection databank from the Delft University of Technology available through the following series of 6 reports \cite{IID1Rep,IID2Rep,IID3Rep,IID4Rep,IID5Rep,IID6Rep}.

In this article, it is proposed to address the optimization under uncertainty of the weight of ring-stiffened cylindrical shells representative of those used in submarine pressure hulls.
The geometric imperfection of a single bay is considered as random in the shape of a few selected critical buckling modes as previously addressed in \citep{Dubourg2008ASRANet}.
The uncertainty in material properties and thicknesses is accounted for by means of a random variable approach.
This uncertainty is therefore not modeled as space-variant random fields as reported in some of the previously cited works.
The parameters of the random models are chosen in accordance to requirements imposed by standards or general recommendations, if available, or arbitrarily set up to prescribed values.
The limit loads are accurately assessed by FE with a non-incremental non-iterative method known as the asymptotic numerical method (ANM).
A nonlinear elastic behavior of the material is assumed in the FE analysis with random material properties (Young's modulus, yield strength and ultimate strength).
An approach based on semi-numerical models is also proposed as an alternative for computing the limit loads.
The weight optimization of the single bay design is carried out under the constraint that the failure probability of the imperfect shell \wrt buckling is kept lower than a prescribed target failure probability.
The proposed approach is therefore a reliability-based design optimization (RBDO) as known in the literature, see \eg \citep{Tsompanakis2008}.
Since the limit loads are predicted by means of a costly-to-evaluate FE analysis, the following constraints are taken into account:
\begin{itemize}
	\item the proposed approach shall be applicable within a few hundred runs of the FE model only: this implies resorting to metamodeling techniques. A recourse to an adaptive Kriging method developed in a so-called augmented space is made \citep{Dubourg2011SMO} .
	\item the probability of failure shall be evaluated precisely at each step of the optimization algorithm and it is likely to be small (\eg less than $10^{-8}$ for the forecast applications). Thus subset simulation \citep{Au2001} is used.
\end{itemize} 
\par

This paper is organized as follows.
\secref{sec:nl_stability_analysis} first introduces the fundamental concepts of nonlinear stability analysis in structural mechanics.
It also defines the assumptions that are accounted for in the subsequent FE-based probabilistic buckling analyses.
FE-based reliability analyses are known to be too-computationally-demanding for being applicable to industrial concerns.
In this respect, \secref{sec:surrogate_based_reliability_analysis} exposes the key concepts of the Kriging surrogate-based strategy.
In \secref{sec:submarine_hull}, RBDO is applied to a single bay of a submarine pressure hull with random geometric imperfections.
This probabilistic design philosophy is opposed to the state-of-the-art worst case approach.
\par

\section{\label{sec:nl_stability_analysis}Elements of shell nonlinear stability analysis}

Buckling is a structural instability phenomenon triggered by some excessive load that needs to be identified.
This load will be referred to as the critical buckling (or collapse) load in the sequel.
In practice it is determined by applying a so-called load proportionality factor (LPF) $\lambda$ which is initialized to zero and then incrementally increased until collapse is observed.
\par

\subsection{Problem formulation}

In continuum mechanics, the equilibrium state of a conservative mechanical system is characterized by a zero elementary variation of its total potential energy denoted by $E_{\rm t}$.
This fundamental principle leads to the establishment of the following so-called variational formulation of equilibrium states:
\begin{equation} \label{eq:zero_variation_of_energy}
	\delta E_{\rm t} = E_{{\rm t},\,\ve{u}}\,(\lambda,\,\ve{u})\,\delta\ve{u} = 0,
\end{equation}
where $\ve{u}$ denotes any admissible displacement of the structure and $E_{{\rm t},\,\ve{u}}$ is the first-order functional derivative of the total potential energy that depends on the LPF $\lambda$.
The infinite set of values of $\lambda$ and $\ve{u}$ satisfying the latter equation is known as the equilibrium path of the structure.
This path is usually constructed incrementally from a known initial state $\lambda^{(0)},\,\ve{u}^{(0)}$, \eg the reference state of the unloaded structure for which $\lambda = 0$.
\par

For stable structures, the only state of interest corresponds to a unit value of the LPF $\lambda$.
Unstable (resp. stable) equilibrium states are characterized by a negative (resp. strictly positive) second-order functional derivative of the total potential energy $E_{{\rm t},\,\ve{u}\,\ve{u}}$, meaning that they correspond to local maxima (resp. minima) of this energy.
\par

\begin{figure}
	\centering
	\includegraphics[width=.55\columnwidth]{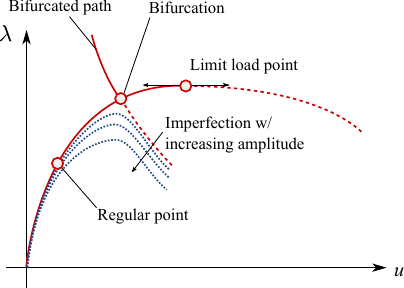}
	\caption{Equilibrium paths and stability}
	\label{fig:stability_equilibrium_path}
\end{figure}

There exists basically two kinds of instabilities, both potentially leading to buckling and/or premature plastic collapse: bifurcation points and limit points, see \figref{fig:stability_equilibrium_path} for illustration.
Regarding bifurcation points, the structure may lose its stability along the equilibrium path, resulting in sudden and large displacements which often lead to collapse.
Regarding limit points, this occurs when the structure is no longer able to withstand loads due to nonlinear geometrical and/or material effects.
For many structures including shells, these two kinds of points generally interact in a joint manner, one triggering the other and conversely.
\par

Practical detection of these instabilities is a non-trivial task and it involves the resolution of a perturbed equilibrium problem along with the resolution of \eqref{eq:zero_variation_of_energy}.
The present study focuses on imperfect structures which are commonly assumed to fail at their limit load.
Indeed, \citet{Koiter1945} showed that the presence of initial imperfections in the structure\footnote{The initial study was carried out on cylinders under a compressive axial load with modal imperfections, but this generalizes to other imperfect structures.} smooths the equilibrium paths.
As a result, imperfect structures regularly feature smooth limit load points rather than sharp bifurcation points which are only observed on perfect structures (see \figref{fig:stability_equilibrium_path}).
Therefore, in the sequel, the detection of singular points along the equilibrium path will be restricted to limit load points characterized by horizontal tangents of the equilibrium path.
\par

\subsection{General formulation of the static equilibrium equations}

In a static analysis, the total potential energy of a structure of volume $\calv$ is given by:
\begin{equation}
	E_{\rm t}(\lambda,\,\ve{u}) = \int_{\calv} W_{\rm int}(\ve{\varepsilon})\,\di{v} - \lambda\,W_{\rm ext}(\ve{u}),
\end{equation}
where $W_{\rm int}$ is the strain energy density in the structure, $W_{\rm ext}$ is the work of external forces and $\di{v}$ is the infinitesimal volume.
$\ve{\varepsilon}$ stands for the Green-Lagrange strain tensor which is defined as:
\begin{equation}
	\ve{\varepsilon} = \ve{\varepsilon}(\ve{u}) = \underbrace{\frac{1}{2}\,\left(\nabla \, \ve{u} + \nabla\tr \ve{u}\right)}_{\displaystyle\ve{\varepsilon}_{\rm l}(\ve{u})} + \underbrace{\frac{1}{2}\,\nabla \, \ve{u}\,\nabla\tr \ve{u}}_{\displaystyle\ve{\varepsilon}_{\rm nl}(\ve{u},\,\ve{u})},
\end{equation}
where $\ve{\varepsilon}_{\rm l}(\ve{u})$ (resp. $\ve{\varepsilon}_{\rm nl}(\ve{u},\,\ve{u})$) denotes the linear (resp. symmetric quadratic bilinear) term of $\ve{\varepsilon}$ and $\nabla$ the gradient operator.
Assuming linear elasticity, the strain energy density $W_{\rm int}$ reduces to the following quadratic form:
\begin{equation}
	W_{\rm int}(\ve{\varepsilon}) = \frac{1}{2} \, \ve{\varepsilon}:\ma{D}:\ve{\varepsilon},
\end{equation}
where $\ma{D}$ is the elasticity tensor of the material and the symbol $:$ denotes the double contraction of tensors.
\par

Introducing the second Piola-Kirchhoff stress tensor $\ve{S} = \ma{D}:\ve{\varepsilon}$, the variational formulation of the equilibrium equation rewrites as the following set of equations:
\begin{equation} \label{eq:structural_equilibrium}
	\left\{\begin{array}{rcl}
		\delta E_{\rm t} & = & \displaystyle\int_{\calv} \ma{S}:\delta \ve{\varepsilon}\,\di{v} - \lambda\,W_{\rm ext}(\delta \ve{u}) = 0 \\[5pt]
		\ma{S} & = & \ma{D}:\ve{\varepsilon}
	\end{array}\right.,
\end{equation}
where $\delta\ve{\varepsilon} = \ve{\varepsilon}_{\rm l}(\delta\ve{u}) + \ve{\varepsilon}_{\rm nl}(\ve{u},\,\delta\ve{u})$.
\par

\subsection{The asymptotic numerical method}

The nonlinear problem in \eqref{eq:structural_equilibrium} is usually solved by means of so-called incremental iterative methods such as the Newton-Raphson algorithm.
The present work is based on an original alternative known as the asymptotic numerical method (ANM) proposed by \citet{Damil1990} and \citet{Cochelin1994CS}.\par

\subsubsection{The idea}

It is first proposed to rewrite the nonlinear problem in \eqref{eq:structural_equilibrium} into the following convenient quadratic form:
\begin{equation} \label{eq:quad_bilin}
	\ve{R}(\ve{Y},\,\lambda) = L(\ve{Y}) + Q(\ve{Y},\,\ve{Y}) - \lambda\,\ve{F} = 0,
\end{equation}
where $\ve{R}$ is a vector of residuals, $L$ is a linear operator, $Q$ is a bilinear quadratic operator, $\ve{F}$ is a known vector and $\ve{Y}\tr = (\ve{u}\tr, \ma{S}\tr)$ groups the unknowns of the problem.
\par

A key idea of the ANM then consists in expanding the unknowns $\ve{Y}$ and $\lambda$ over a unique path parameter denoted by $a$ in the form of the following polynomial series expansions:
\begin{equation} \label{eq:ANM_expansion}
	\left\{\begin{array}{rcl}
		\ve{Y}(a) & = & \ve{Y}_0 + a\,\ve{Y}_1 + a^2\,\ve{Y}_2 + \ldots + a^N\,\ve{Y}_N \\
		\lambda(a) & = & \lambda_0 + a\,\lambda_1 + a^2\,\lambda_2 + \ldots + a^N\,\lambda_N
	\end{array}\right.,
\end{equation}
where $(\ve{Y}_0,\,\lambda_0)$ describes the initial state of the system, supposedly known.
In this study, the polynomial expansions are truncated after $N = 30$ terms.
\par

Introducing these expansions into \eqref{eq:quad_bilin} and grouping the terms with the same power of $a$ then yield the following succession of linear systems for orders $p = 1\enu N$:
\begin{equation} \label{eq:linear_systems}
	\left\{\begin{array}{rcl}
		L_{\rm t}(\ve{Y}_1) & = & \lambda_1\,\ve{F} \\
		L_{\rm t}(\ve{Y}_2) & = & \lambda_2\,\ve{F} - Q(\ve{Y}_1,\,\ve{Y}_1) \\
		& \vdots & \\
		L_{\rm t}(\ve{Y}_N) & = & \lambda_N\,\ve{F} - \sum\limits_{p=1}^{N-1} Q(\ve{Y}_p,\,\ve{Y}_{N - p})
	\end{array}\right.,
\end{equation}
where $L_{\rm t}(\bullet) = L(\bullet) + 2\,Q(\ve{Y}_0,\,\bullet)$ is the tangent operator, which is the same at all orders.
\par

At this stage, the problem involves one more unknown than the number of available equations, namely the parameter $a$.
Similarly to a classical incremental iterative method, the ANM uses a pseudo arc-length technique by setting:
\begin{equation}
	a = \left(\ve{Y} - \ve{Y}_0\right)\tr \ve{Y}_1 + \left(\lambda - \lambda_0\right)\,\lambda_1,
\end{equation}
which completes the system of equations in \eqref{eq:linear_systems}.
\par

Hence, it can be seen that the initial nonlinear problem in \eqref{eq:quad_bilin} has been genuinely transformed into a set of $N$ linear systems by rejecting all nonlinearities to the right-hand side of \eqref{eq:linear_systems}.
In addition, the $N$ linear systems composing \eqref{eq:linear_systems} feature a single linear operator $L_t$ which is the same at all orders.
When switching to the discrete form of the problem (by means of a classical FE displacement formulation), the resolution of the $N$ linear systems requires only one decomposition of the tangent stiffness matrix $\ma{K}_{\rm t}$ which is the discrete counterpart of $L_{\rm t}$.
This latter remark makes the ANM very efficient as the tangent stiffness matrix $\ma{K}_{\rm t}$ is large in practice.
\par

Eventually, the ANM provides a continuous representation of the equilibrium path for any value of $\lambda$ thanks to the series expansion in $a$.
This is an interesting property with respect to the incremental iterative methods that need to solve the problem for each value of $\lambda$.
\par

\subsubsection{Validity of the expansion}

Due to the use of finite expansions in \eqref{eq:ANM_expansion}, the solution becomes invalid for large values of $a$.
Thus, it is proposed to truncate the solution below a maximum value of $a$ denoted by $a_{\max}$.
This maximum value is based on a study of the norm of the residual $\ve{R}(a) = \ve{R}(\ve{Y}(a),\,\lambda(a))$.
\par

\citet{Cochelin1994CS} proved that it is reasonable to approximate this quantity by the norm of the first omitted term in the expansion, so that:
\begin{equation}
	\left\|\ve{R}(a)\right\| \approx \left\|a^{N+1}\,\ve{R}_{N+1}\right\|.
\end{equation}
Based on this approximation, \citeauthor{Cochelin1994CS} then came up with the following expression for $a_{\max}$:
\begin{equation}
	a_{\max} = \left(\epsilon\frac{\left\|\ve{F}\right\|}{\left\|\ve{R}_{N+1}\right\|}\right)^{\frac{1}{N + 1}},
\end{equation}
where $\epsilon$ is the maximum tolerance on the norm of the residual. This tolerance is usually set equal to a small value (here $10^{-8}$) thanks to the normalization of the residual with respect to the right-hand side $\left\|\ve{F}\right\|$ of \eqref{eq:quad_bilin}.\par

The description of the whole equilibrium path is therefore made piecewise by repeating the procedure incrementally, \ie by resetting the initial state of the system $(\ve{Y}_0,\,\lambda_0)$ to $(\ve{Y}(a_{\max}),\,\lambda(a_{\max}))$. It is worth pointing out that the ANM remains more computationally efficient than its incremental iterative counterparts because it is incremental only. Indeed, incremental iterative methods need to iterate within the increments in order to remain on the equilibrium path, which involves expensive decompositions of the tangent stiffness matrix during the iterative process.

\subsubsection{Determination of the limit load carrying capacity} \label{sec:limit_detection}

The determination of the limit-load carrying capacity exploits the parametric approximation of the load proportionality factor.
Indeed, limit load points are characterized by an horizontal tangent on the equilibrium path thus meaning that the derivative of the load proportionality factor with respect to $a$ equals zero at the critical limit load.
Hence, the limit LPF is defined as $\lambda_{\rm limit} = \lambda(a_{\rm limit})$ where:
\begin{equation}
	a_{\rm limit} = \min \acc{a \in \bra{0;\;a_{\max}}: \frac{\di{\lambda}}{\di{a}} = 0}.
\end{equation}
Thanks to the chosen polynomial series expansion for the LPF, determining the limit load simply consists in finding the roots of a polynomial of order $\prt{N-1}$ and retaining the lowest positive root that is less than $a_{\rm max}$, provided it exists.
\par

\subsubsection{Sources of nonlinearity} \label{sec:sources_nonlin}

In the present work, the ANM is applied to geometric nonlinearities extended to large rotations based on the work of \citet{Zahrouni1999}.
It can be shown that the corresponding equilibrium equations under such an assumption conveniently fit in the quadratic formulation of the ANM given in \eqref{eq:quad_bilin}.
Two additional sources of nonlinearity are explicitly accounted for within the ANM.
\par

The first source of nonlinearity is due to follower forces resulting from the hydrostatic pressure exerted on the shell.
Accounting for the specific effects of follower forces in a buckling analysis could be of utmost importance for some structural components such as the single-bay of the submarine pressure hull with an overall geometrical imperfection studied in \secref{sec:submarine_hull}.
From a computational viewpoint, this additional assumption introduces a dependence of the virtual work of external forces on the LPF \citep[see \eg{}][pp. 81--86]{Noirfalise2009PhD}.
This results in additional terms appearing in the right-hand side forces of \eqref{eq:linear_systems}, on the one hand, and a nonsymmetric tangent operator $L_{\rm t}$ (nonsymmetric tangent stiffness matrix $\ma{K}_{\rm t}$ in the matrix formulation), on the other hand.
\par

The second source of nonlinearity accounted for in this work is due to the assumption of a nonlinear behavior of the constitutive material of the shell.
A nonlinear elastic Ramberg-Osgood constitutive law characterized by the following stress-strain relationship is considered within the ANM \citep{Zahrouni1998}:
%\begin{equation}
%    \varepsilon = \frac{\sigma}{E} + \alpha\,\frac{\sigma}{E}\,\left(\frac{\left|\sigma\right|}{\sigma_{\rm y}}\right)^{n-1},
%\end{equation}
\begin{equation}
	E \ve{\varepsilon} = \left(1+\nu\right) \ve{S}^{\rm d} - \left(1-2\nu\right) P \Id + \frac{3}{2} \alpha \left(\frac{S_{\rm eq}}{\sigma_{\rm y}}\right)^{n-1} \ve{S}^{\rm d},
\end{equation}
where $E$ is the Young's modulus, $\nu$ is the Poisson's ratio, $\sigma_{\rm y}$ is the yield strength, $P = -\frac{1}{3} \ve{S} : \Id = -\frac{1}{3} \operatorname{tr}\left(\ve{S}\right)$ is the hydrostatic pressure, $\ve{S}^{\rm d} = \ve{S} + P \Id$ is the deviatoric part of the stress tensor $\ve{S}$, $S_{\rm eq} = \sqrt{ \frac{3}{2} \ve{S}^{\rm d} : \ve{S}^{\rm d}}$ is the von Mises equivalent stress, $\alpha$ and $n$ are the two Ramberg-Osgood parameters.
\par
Note that plasticity is not taken into account in the present analysis.
Even though, it is argued that the structure under concern here does not present any significant local unloading until the collapse load of interest is reached.
In such a case, nonlinear elasticity represents a fairly accurate model.
\par

As additional details, it is important to mention that large rotations and shear strains in the thin-walled shell structure of interest are accounted for by means of the shell FE formulation.
The present approach resorts to a three-dimensional seven-parameter shell formulation proposed by \citet{Buchter1994}.
This formulation based on the enhanced assumed strain (EAS) concept disables the usual locking problems featured by shell elements.
However, it introduces another set of nonlinear compatibility equations \citep[see \eg{}][pp. 43--48]{Baguet2001PhD}.
\par

%%%%%%%%%%%%%%%%%%%%%%%%%%%%%%%%%%%%%%%%%%%%%%%%%%%%%%%%%%%%%%%%%%%%%%
\section{\label{sec:surrogate_based_reliability_analysis}RBDO using an adaptive Kriging metamodel}

\subsection{Problem statement}

Reliability-based design optimization (RBDO) aims at obtaining an optimal design which guarantees a chosen reliability level \wrt various performance criteria.
More specifically a cost function $c$ is to be minimized by selecting an optimal set of design parameters denoted by $\ve{d}^*$ while fulfilling probabilistic constraints:
\begin{equation} \label{eq001}
	% Use of the notation b_i(d) for the constraints
	% instead of f_i(d). f_i is used for Kriging regression functions
	\begin{split}
		\ve{d}^* = & \arg \min\limits_{\ve{d}\in\cald_{\ve{d}}} c\left(\ve{d}\right)\;:\\
		& \left\{\begin{array}{l}
			b_i\left(\ve{d}\right) \leq 0,\;i = 1\enu n_c \\[2pt]
			\Prob{g_l\left(\ve{X}\left(\ve{d}\right)\right) \leq 0} \leq p_{f\,l}^0,\;l = 1\enu n_p
		\end{array}\right.
	\end{split}
\end{equation}
In this equation $\cald_{\ve{d}}$ is the design space, $\{b_i\left(\ve{d}\right) \leq 0, i= 1 \enu n_c\}$ are deterministic feasibility constraints on the parameters (also called soft constraints such as bounds) and $\{\Prob{g_l\left(\ve{X} \left(\ve{d}\right)\right) \leq 0} \leq p_{f\,l}^0, l = 1 \enu n_p\}$ are the reliability constraints meaning that the retained design $\ve{d}^*$ should lead to failure probabilities smaller than the requirements $\{p_{f\,l}^0, l = 1 \enu n_p\}$.
The notation $\Ve{X}(\ve{d})$ is used to emphasize that the input random vector modeling the uncertainty (of prescribed PDF $f_{\Ve{X}} (\ve{x} | \ve{d})$) contains two types of variables:
\begin{itemize}
	\item[-] variables whose mean values are the design parameters $\ve{d}$: their variability correspond to unavoidable scattering in the manufacturing of the system;
	\item[-] environmental parameters (\eg loads) and possibly material properties that are uncertain in nature.
\end{itemize}

As argued in the introduction, the complex structural behaviors that are to be addressed by the proposed approach cannot usually be dealt with by approximate reliability methods (such as FORM/SORM) due to the complex shapes of the limit-state surfaces $\acc{g_l\prt{\Ve{X} \prt {\ve{d}}} = 0,\,l = 1 \enu n_p}$.
In contrast, simulation methods such as Monte Carlo or more advanced approaches such as subset simulation \citep{Au2001} are unaffordable when using FE models.
This naturally leads to introducing metamodels for both computing the failure probabilities and optimizing the system.
In the present paper, Kriging surrogates \citep{Santner2003} are used in an augmented space which enables solving the two issues in one single step.
\par

\subsection{Kriging surrogates: a summary}

Let us consider a single performance function $g(\ve{x}),\; \ve{x} \in \cald_{\ve{x}} \subseteq \Rr^n$.
Kriging is a statistical learning technique that comes from geostatistics and that is now used in computer experiments \citep{Sacks1989}.
A Kriging metamodel is an analytical function $\tilde{g}$ that is inexpensive to evaluate (\wrt a performance function that involves running a large FE model) and that may be built from an experimental design denoted by $\calx = \{\ve{x}_i \in \cald_{\ve{x}},\,i = 1 \enu m\}$ and the associate performances gathered in $\ve{y} = \{g(\ve{x}_i),\,i = 1 \enu m\}$.
\par

Kriging assumes that the function $g$ of interest is a realization of a Gaussian process denoted by $Y(\ve{x}),\,\ve{x}\in\cald_{\ve{x}}$ which is defined as follows:
\begin{equation} \label{eq:029}
	Y\left(\ve{x}\right) = \ve{f}\left(\ve{x}\right)\tr\,\ve{a} + Z\left(\ve{x}\right)
\end{equation} 
In this equation $\ve{f}\left(\ve{x}\right)\tr\,\ve{a}$ is the mean of the process, which is represented by a set of basis functions $\acc{f_i,\,i = 1 \enu P} $ (\eg polynomial functions) and $Z\left(\ve{x}\right),\,\ve{x}\in\cald_{\ve{x}}$ is a stationary zero-mean Gaussian process with variance $\sigma_Y^2$ and autocorrelation function:
\begin{equation} \label{eq:030}
	C_{YY}\left(\ve{x},\,\ve{x}'\right) = \sigma_Y^2\,R\left(\ve{x}-\ve{x}'\, ,\,\ve{\theta}\right), \quad \left(\ve{x},\,\ve{x}'\right) \in \cald_{\ve{X}}\times\cald_{\ve{X}}
\end{equation}
In the above equation $\ve{\theta}$ gathers all the parameters defining $C_{YY}$.
In practice, square exponential models are generally postulated \citep{Lophaven2002}:
\begin{equation} \label{eq:031}
	R\left(\ve{x}-\ve{x}',\ve{\theta}\right) = \exp\left[\sum\limits_{k=1}^n - \left(\frac{x_k-x_k'}{\theta_k}\right)^2\right]
\end{equation}
The Kriging estimator at a test point $\ve{x} \in \cald_{\ve{\ve{x} }}$ is obtained by the best linear unbiased estimator (BLUE) which is the linear combination of the observations (\ie the points in $\calx$) that provides a minimal variance.
By construction the Kriging estimator is a Gaussian random variable $\widehat{Y}\left(\ve{x}\right) \sim \caln\left(\mu_{\widehat{Y}} \left(\ve{x}\right),\,\sigma_{\widehat{Y}}\left(\ve{x}\right)\right)$ whose mean value is the metamodel of interest:
\begin{equation} \label{eq:032}
	\tilde{g} (\ve{x} ) \equiv \mu_{\widehat{Y}}\left(\ve{x}\right) = \ve{f}\left(\ve{x}\right)\tr\,\hat{\ve{a}} + \ve{r}\left(\ve{x}\right)\tr{\ma{R}}^{-1}\left(\ve{y} - \ma{F}\,\hat{\ve{a}}\right)
\end{equation}
In this equation the following notations $\ve{r}$, $\ma{R}$ et $\ma{F}$ are used:
\begin{eqnarray}
	r_i(\ve{x}) & = & R\left(\ve{x} - \ve{x}_i,\,\ve{\theta}\right), \quad i = 1 \enu m \\
	R_{ij} & = & R\left(\ve{x}_i-\ve{x}_j,\,\ve{\theta}\right), \quad i,\,j = 1 \enu m \label{eq:034} \\
	F_{ij} & = & f_j\left(\ve{x}_i\right), \quad i = 1 \enu m,\,j = 1 \enu P
\end{eqnarray}
\par

The Kriging variance $\sigma_{\widehat{Y}} (\ve{x})$ whose expression can be found in \citet{Santner2003} reflects the precision of the estimator.
It is a measure of the epistemic uncertainty that comes from the limited information on $g$ gathered in the observations $\acc{\calx, \ve{y}}$.
A large value of $\sigma_{\widehat{Y}} (\ve{x})$ means that the metamodel cannot be trusted at point $\ve{x}$ whereas a small value guarantees a good accuracy of the metamodel at this point.
\par

Finally the parameters $\sigma_Y^2,\,\ve{a},\,\ve{\theta}$ are estimated by means of the maximum likelihood principle using the observations in $\ve{y}$ \citep[see \eg][for more details]{Marrel2008,Dubourg2011PhD} or by cross-validation techniques \citep[see \eg][]{Bachoc2013}.
\par

\subsection{Augmented space}

The RBDO problem may be solved through an iterative optimization algorithm in which the failure probability associated to the current values of
the design parameters $\ve{d}_k$ shall be computed for each iteration $k$.
In a naive approach a new metamodel could be built at each iteration consistently with the domain of variation of the current probabilistic model $\Ve{X} (\ve{d}_k)$.
In order to avoid such a computational burden \citet{Dubourg2011SMO} propose to build a single global Kriging surrogate in an augmented space that ``mixes'' both the (supposedly bounded) design space $\cald_{\ve{d}}$ (of volume $\calv_{\cald_{\ve{d}}}$) and the randomness in $\Ve{X} (\ve{d})$.
This corresponds to defining an augmented distribution $h(\ve{v})$:
\begin{equation}
	h\left(\ve{v}\right) = \int_{\cald_{\ve{d}}} f_{\ve{X}}\left(\ve{x} \left| \ve{d} \right.\right)\,\pi\left(\ve{d}\right)\,\di{\ve{d}}
	%  = \frac{1}{\calv_{\cald_{\ve{d}}}} \int_{\cald_{\ve{d}}}
	%  f_{\ve{X}}\left(\ve{x} \left| \ve{d} \right.\right)\,\di{\ve{d}} 
\end{equation}
where $\pi\left(\ve{d}\right)$ denotes a prior probability distribution on the design space, which is usually a uniform distribution when no specific design is preferred a priori.
In this case one gets:
\begin{equation} 
	h\left(\ve{v}\right)\propto \int_{\cald_{\ve{d}}} f_{\ve{X}}\left(\ve{x} \left| \ve{d} \right.\right)\,\di{\ve{d}}
\end{equation}
The space on which the global Kriging surrogate is built is constructed as a confidence domain $\cald_{\ve{V}}$ that is sufficiently large to allow an accurate reliability estimation whatever the current probabilistic model $\Ve{X} (\ve{d}_k)$.
In the case when the variables associated with the design parameters are independent, bounds may be computed for each single component $X^{(j)}(d_k^{(j)})$ and the confidence domain $\cald_{\ve{V}}$ is the hyperrectangle obtained by tensor products of these ranges \citep[see][for more details]{Dubourg2011SMO,Dubourg2011PhD}.
\par

\subsection{Adaptive Kriging}

\begin{figure}
	\centering
	\includegraphics[width=.7\columnwidth,clip=true,trim=0 30 0 40]{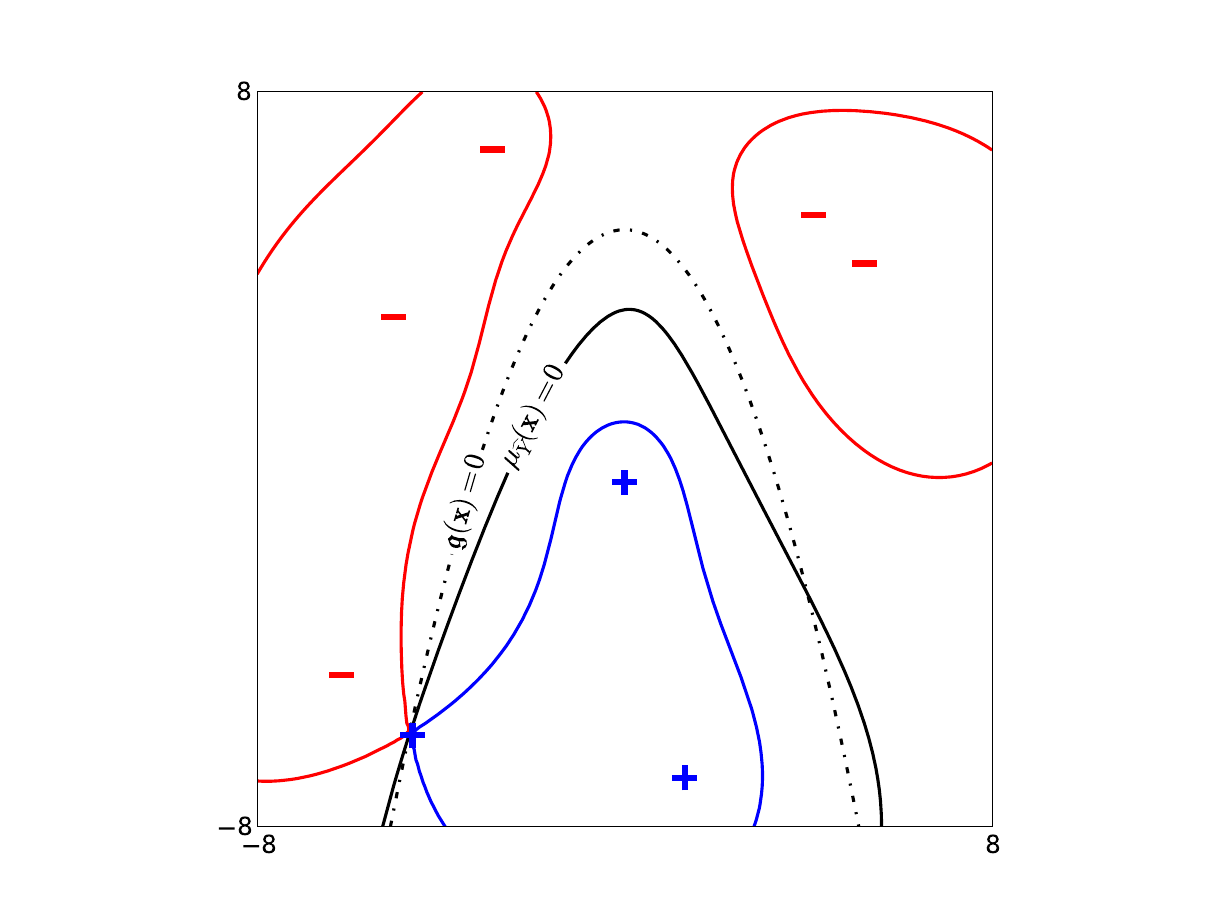}
	\caption{Example of a true limit-state surface, its Kriging surrogate and the bounds of the margin of uncertainty.}
	\label{fig:001}
\end{figure}

As shown in the previous paragraph the Kriging surrogate shall be precise enough in order to compute the failure probability for various values of the current design parameters $\ve{d}_k$.
The Kriging variance provides a measure of ``how good the surrogate is'' at each point.
However in a reliability analysis, the quantity of interest is the limit-state surface \ie the contour $\acc{g(\ve{x}) =0}$. Hence from the metamodel $\tilde{g} = \mu_{\widehat{Y}}$ and the Kriging variance \citet{Dubourg2011SMO} define a margin of uncertainty in the vicinity of this contour:
\begin{equation} \label{eq:039}
	\mathfrak{M} = \acc{\ve{x}\,:\,
		-k\,\sigma_{{\widehat{Y}}}\left(\ve{x}\right) \leq
		\mu_{\widehat{Y}}\left(\ve{x}\right) \leq
		+k\,\sigma_{{\widehat{Y}}}\left(\ve{x}\right)}
\end{equation}
where $k$ is a ``number of standard deviations'' of the epistemic uncertainty.
Heuristically it is a domain in which there is a high probability (\eg 95\% for $k=1.96$) that the true limit-state surface lies in, up to the validity of Kriging assumptions (see \figref{fig:001}).
\par

The use of this margin (which is a subset of $\cald_{\Ve{x}}$) is twofold:
\begin{itemize}
	\item[-] Its boundaries define two domains which yield two estimates of the probability of failure for the current domain. The closeness of these two values denoted by $p_f^-$ and $p_f^+$ will be an indicator of the precision of the reliability estimates and in most cases the true value of $p_f$ lies between these bounds although there is no formal proof (see also \citet{Schobi2016} for more details).
	\item[-] A probabilistic classification function is defined by:
	\begin{equation} \label{eq:038}
		\pi(\ve{x},t) = \calp\bra{\hat Y(\ve{x}) \le t} = \Phi\prt{\frac{t-\mu_{\widehat{Y}}\left(\ve{x}\right)}{\sigma_{\widehat{Y}}\left(\ve{x}\right)}}
	\end{equation}
	Note that in this equation, $\calp\bra{\bullet}$ denotes the Gaussian probability measure associated with the epistemic uncertainty of Kriging and not the aleatoric uncertainty in $\Ve{X} $ (the latter being denoted by $\Pp\bra{\bullet}$).
\end{itemize}
The probability that a point $\ve{x}$ belongs to the margin of uncertainty $ \mathfrak{M}$ is readily obtained by:
\begin{equation} \label{eq:040}
	\calc({\ve{x} }) = \calp\bra{\widehat{Y}(\ve{x})\in \mathfrak{M}} = \pi(\ve{x},\,k\,\sigma_{\widehat{Y}}(\ve{x}) ) - \pi(\ve{x},\,-k\,\sigma_{\widehat{Y}}(\ve{x}))
\end{equation}
This quantity $\calc(\ve{x})$ is interpreted as a sampling PDF whose maximal values correspond to being in the margin of uncertainty, \ie close to the zero-value of the limit-state function and in regions where there is a lack of accuracy in the surrogate.
By using this PDF (defined up to its normalizing constant) and a Markov chain Monte Carlo sampling technique such as slice sampling \citep{Neal2003}, one can draw a large set of points that lie in the margin $\mathfrak{M}$, \ie that are potential candidates for enriching the experimental design $\calx$.
A technique of $K$-means clustering helps reducing this number of points to a set of limited size (say $K_{\rm enrich} =$~ 5 to 10) which are added to the experimental design.
A new Kriging metamodel is then built from this updated experimental design.
\par

In the present approach, the Kriging surrogate is refined by repeating the above procedure until the log-ratio of the probabilities of failure obtained from the bounds of $\mathfrak{M}$ is smaller than an admissible value, \ie $\log(p_f^+/p_f^-) \leq \varepsilon_{p_f}$.
\par

\subsection{Proposed RBDO algorithm}

The proposed RBDO algorithm relies upon the gradient-based Polak-He algorithm \citep[see][]{Polak1997,Dubourg2011PhD}.
Two types of convergence criteria shall be checked for:
\begin{itemize}
	\item[-] the Kriging surrogate shall be accurate enough to evaluate the failure probability in each iteration, as previously described;
	\item[-] the optimal design of the structure should be converged.
\end{itemize}

As a summary of the above sections, the proposed RBDO algorithms reads as follows:
\begin{enumerate}
	\item {\bf Initialization}: Define the ranges of the design parameters and an initial value $\ve{d}_0$. Define the conditional probabilistic model   $\ve{X}(\ve{d})$. Compute the confidence domain $\cald_{\ve{V}}$. Define an initial experimental design $\calx_0$ on this domain and build the initial Kriging surrogate.
	\item {\bf Optimization iteration $k$}
	\begin{itemize}
		\item[$\bullet$] Estimate the current failure probability $p_f(\ve{d_k})$ and associated bounds obtained from the Kriging margin of uncertainty. Possibly add new points to $\calx$ and refine the Kriging surrogate until the error on $p_f(\ve{d_k})$ is considered acceptable.
		\item[$\bullet$] Improve the current design by one step of the gradient-based optimization algorithm (\ie $\ve{d}_{k+1}= \text{PolakHe}(\ve{d}_k)$). This requires the evaluation of the gradient of $p_f$ w.r.t. $\ve{d}$.
		\item [$\bullet$] Check the convergence of the optimization.
	\end{itemize}
	\item {\bf Check} the reliability constraints associated with the final design. A Kriging-based importance sampling scheme proposed by \citet{DubourgPEM2011} is used for this purpose. This latter technique can also be used for the analytical computation of sensitivity measures that are used in the gradient-based optimization algorithm, see \citep{Dubourg2014SS}.
	
\end{enumerate}

%%%%%%%%%%%%%%%%%%%%%%%%%%%%%%%%%%%%%%%%%%%%%%%%%%%%%%%%%%%%%%%%%%%%%%
\section{\label{sec:submarine_hull}Design optimization of a submarine pressure hull}

Submarine pressure hulls are mainly composed of structural components such as ring-stiffened cylinders, cones, elliptical or spherical ends, internal diaphragms, bulkheads and deep frames.
At a diving depth $I$, these structures are subjected to an external hydrostatic pressure $p = \rho_{\rm water}\,g\,I$ where $\rho_{\rm water}$ is the sea water density (set here equal to $1,000$~kg/m$^3$) and $g \approx 10$~m/s$^2$ is the gravitational constant.
Such a loading induces a compression stress state that is mostly membrane dominated.
Buckling therefore constitutes a critical failure mode for submarines.
\par

The design practice is usually based on specific standards and design codes such as the British Standard 5500 (\citeauthor{BS5500}) or the more recent Eurocode 3, possibly along with finite-element-based simulations.
It often makes use of long-term-experience-based safety factors at various design stages, which eventually implies an unknown degree of conservatism.
Hence, structural reliability methods reveal a promising tool for investigating the safety margins attached to the current submarine design practices \citep[see \eg][]{Faulkner1990,Pegg1995,Groen1996,Bourinet2000IASSIACM}.
\par

Another major challenge for the designer consists in finding an optimal ratio between the weight of the resistant structure and the buoyancy of the submersible. The latter point falls under the reliability-based design optimization (RBDO) formulation.
The work presented in the sequel is based on preliminary studies published by \citet{Dubourg2008ASRANet,Dubourg2011ICASP}.
\par

\subsection{The single bay reference structure}

\subsubsection{A ring-stiffened shell cylinder}

\begin{figure}
	\centering
	\includegraphics[width=.65\columnwidth]{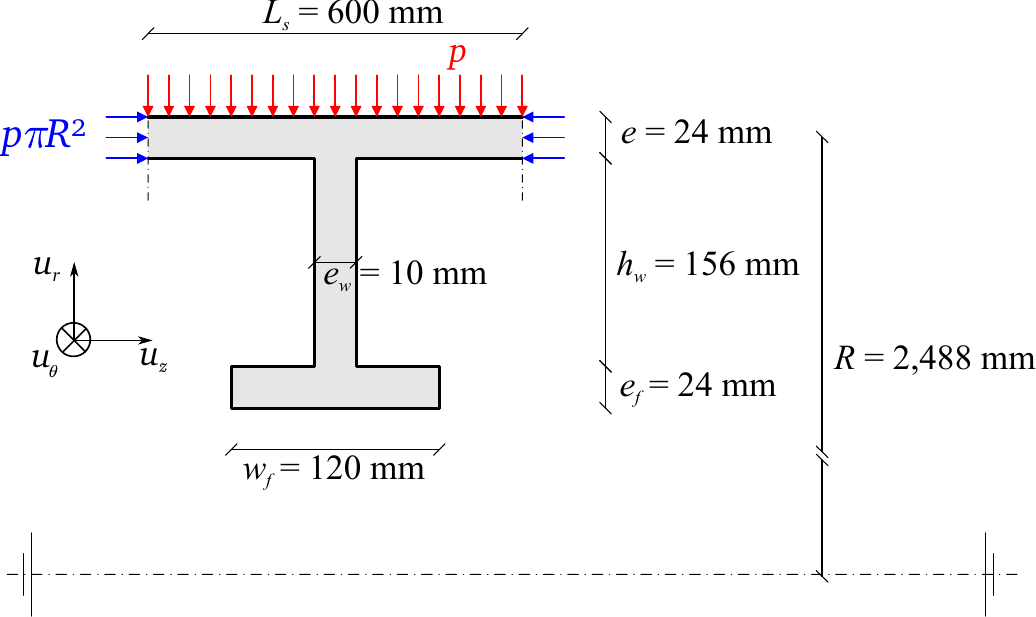}
	\caption{Single bay reference structure and initial design.}
	\label{fig:ring_stiffened_cylinder}
\end{figure}

The present study does not consider the submarine pressure hull as a whole.
It focuses instead on a single bay reference structure consisting of a shell cylinder with a single inner T-section ring stiffener and whose length is equal to the stiffener spacing.
The dimensions of this elementary structure are shown in \figref{fig:ring_stiffened_cylinder}.
In the following, the outer cylinder is referred to as the shell plating, and the web (resp. the flange) designates the vertical (resp. horizontal) part of the T-section ring stiffener.
This simplified model with well-chosen boundary conditions is supposed representative of the behavior of a central bay of a sufficiently long pressure hull compartment (infinite-length in the present study).
\par

\begin{figure*}
	\centering
	\subfigure[Overall mode]{\includegraphics[width=.3\textwidth]{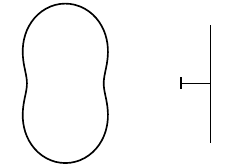}}
	\quad
	\subfigure[Interframe mode]{\includegraphics[width=.3\textwidth]{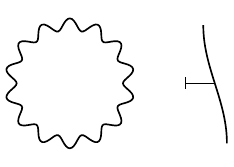}}
	\quad
	\subfigure[Frame tripping mode]{\includegraphics[width=.3\textwidth]{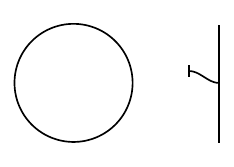}}
	\caption{Schematic representation of the most critical buckling patterns of a ring-stiffened shell cylinder.}
	\label{fig:ring_stiffened_cylinder_buckling_modes}
\end{figure*}

The linear elastic stability analysis of this ring-stiffened shell exhibits some typical buckling patterns.
The three most critical kinds of buckling patterns are known as overall buckling, interframe buckling and frame tripping and they are basically illustrated in \figref{fig:ring_stiffened_cylinder_buckling_modes}.
Actual structures exhibit some unavoidable shape imperfections due to the manufacturing process (mostly cold-bending- and welding-based) and heavy loads connected to the hull (\eg the nuclear reactor).
These initial imperfections may trigger buckling or premature plastic collapse at pressure far below those corresponding to elastic buckling, even if these imperfections are of moderate amplitude due to the stringent tolerances used in fabrication.
\par

Predicting the collapse pressure for any given imperfect geometry is not straightforward though because the structure may feature a considerable degree of interaction between the aforementioned buckling modes.
For solving the buckling problem at hand, the designer may resort to closed-form solutions or other semi-numerical methods available in the codes of practice (\eg the \citeauthor{BS5500}).
Another alternative that is investigated here consists in using an appropriate FE model.
\par

\subsubsection{Modeling of the shape imperfections}

Note that the present analysis is restricted to the effects of overall and interframe shape imperfections.
The collapse due to frame tripping is avoided here by imposing some conservative rules taken from \citeauthor{BS5500} regarding the proportions of the stiffener web and flange during the optimization.
The overall (resp. interframe) radial geometric imperfection is given by:
\begin{eqnarray}
	\zeta_n(z,\,\theta) & = & A_n\,\cos\left(n\,\theta\right), \\
	\zeta_m(z,\,\theta) & = & A_m\,\cos\left(\pi \, \frac{z}{L_{\rm s}}\right)\,\cos\left(m\,\theta\right),
\end{eqnarray}
where $n$ (resp. $m$) is the number of circumferential waves that typically ranges from 2 to 6 (resp. 10 to 20), $A_n$ (resp. $A_m$) denotes the amplitude of the the overall (resp. interframe) radial imperfection, and $0 \leq \theta < 2 \pi$, $0 \leq z \leq L_{\rm s}$.
In the present application, only two modes are considered: $n=2$ and $m=14$.
These two modes correspond to the most critical buckling patterns of the initial design.
A finer study would consist in considering a larger spectrum of imperfections depending on the current design at each iteration of the optimization process.
\par

\subsubsection{Nonlinear finite element model}

It is proposed here to compute the collapse pressure by means of the asymptotic numerical method, accounting for material and geometric nonlinearities.
The steel that constitutes the pressure hull is assumed to follow a nonlinear elastic Ramberg-Osgood constitutive law as described in \secref{sec:sources_nonlin}. Follower forces are taken into account for the hydrostatic pressure field $p$ so that it is always exerted normally with respect to the deformed structure.
\par

\begin{figure*}
	\centering
	\subfigure[Nodes for the elimination of rigid body modes]{
		\includegraphics[width=.4\textwidth]{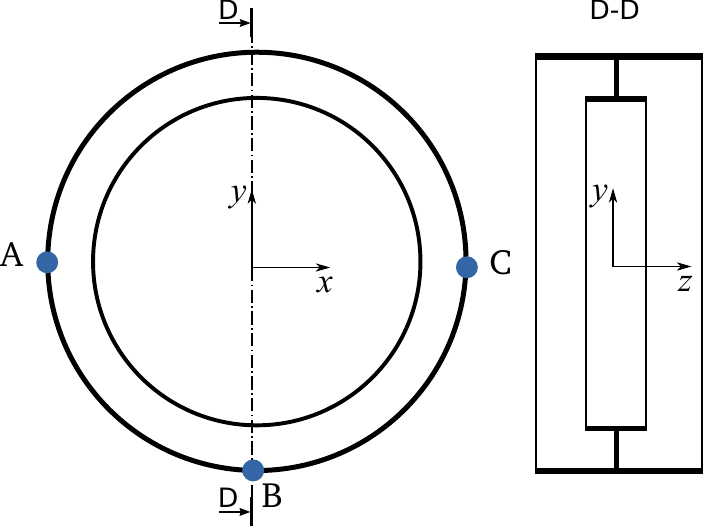}
		\label{fig:ring_stiffened_cylinder_rigid_body_motion}}
	\hfill
	\subfigure[Perfect mesh]{
		\includegraphics[width=.4\textwidth]{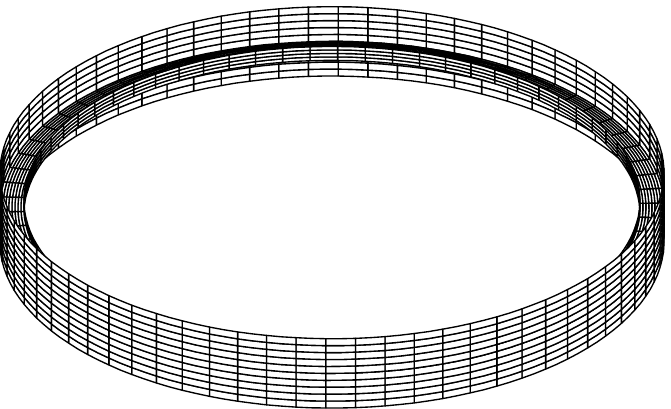}
		\label{fig:mesh}}
	\caption{Finite element modeling of the ring-stiffened shell cylinder.}
\end{figure*}

Rigid body modes are eliminated in three nodes as illustrated in \figref{fig:ring_stiffened_cylinder_rigid_body_motion}:
\begin{itemize}
	\item[-] in A, the three translations are set equal to zero,
	\item[-] in B, the translation along the $z$-axis is set equal to zero,
	\item[-] in C, the translations along the $y$- and $z$-axes are set equal to zero.
\end{itemize}\par

The orthoradial rotations of the two circular ends of the cylinder are set equal to zero in order to fulfill the assumption of repeated adjacent bays.
As an additional hypothesis, these two ends are supposed to remain plane and normal to the $z$-axis during the whole loading process, \ie the nodes of each end cross-section undergo a constant but unknown overall axial displacement.
\par

In addition to the hydrostatic pressure exerted on the outer cylinder, an axial membrane compressive stress of amplitude $p\,\pi\,R^2$ is applied as indicated in \figref{fig:ring_stiffened_cylinder}.
This additional load is due to the hydrostatic pressure exerted on both ends of the pressure hull.
\par

\begin{figure}
	\centering
	\subfigure[Overall imperfection]{
		\includegraphics[width=.3\columnwidth]{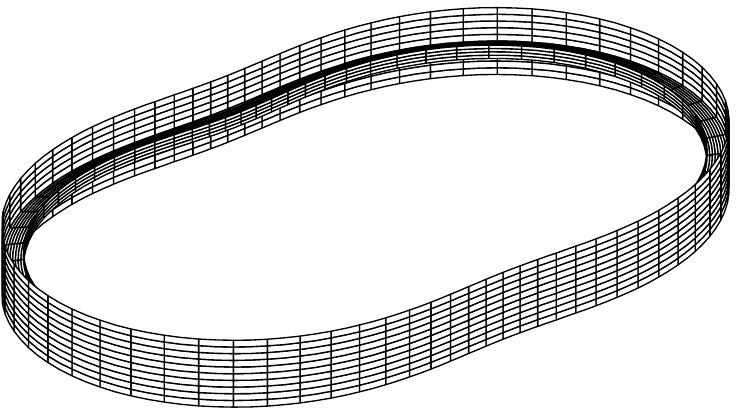}}
	\quad
	\subfigure[Interframe imperfection]{
		\includegraphics[width=.3\columnwidth]{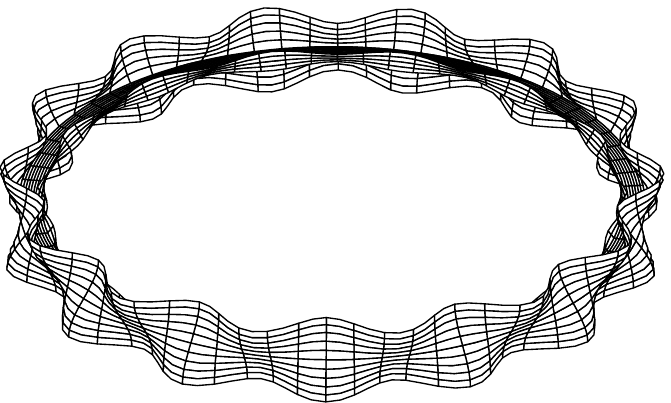}}
	\quad
	\subfigure[Both imperfections]{
		\includegraphics[width=.3\columnwidth]{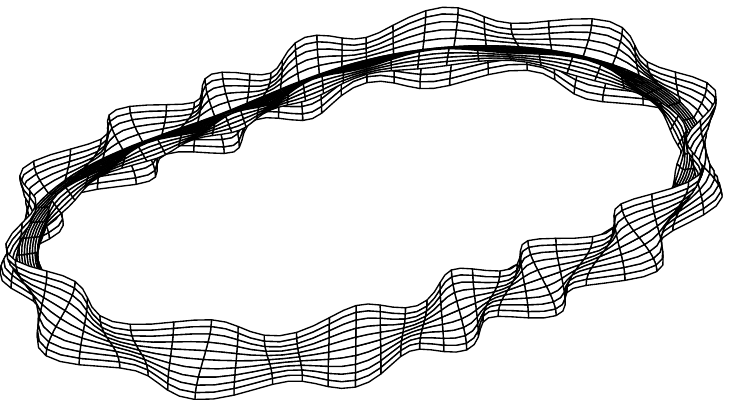}}
	\caption{Ring-stiffened shell cylinder with amplified imperfections.}
	\label{fig:ring_stiffened_cylinder_with_amplified_imperfection}
\end{figure}

The structure is meshed with 1,540 B\"uchter and Ramm 8-node shell elements featuring about 40,000 degrees of freedom: 70 $\times$ 10 elements for the outer cylinder, 70 $\times$ 8 for the web of the stiffener and 70 $\times$ 4 elements for its flange.
The collapse pressure was shown to stabilize for a coarser mesh featuring 15,000 degrees of freedom although it has been raised here in order to accurately represent the highest modal imperfection featuring 14 waves along the circumference, one wave being represented here with 70/14~=~5 elements.
Amplified superpositions of the two geometric imperfections considered here are illustrated in \figref{fig:ring_stiffened_cylinder_with_amplified_imperfection}.
\par

\subsubsection{Semi-numerical model}

In the sequel the designs obtained with the ANM-based FE model are compared with those obtained from approximate semi-numerical solutions available in the submarine pressure hull design codes of practice \citep[see][for a review]{Dubourg2008ASRANet}.
These approximations suppose in general a geometrical imperfection of a given modal shape and, as a consequence, they are not able to account for the possible interactions between buckling modes in case of multimodal (therefore more general) imperfections.
The model for predicting the overall plastic collapse pressure $p_{n\,{\rm pl}}$ is based here on the Bryant formula embedded in the \citeauthor{BS5500}.
The one used for the interframe plastic collapse pressure $p_{m\,{\rm pl}}$ resorts to an interpolated table of numerical solutions derived by the Krylov Shipbuilding Research Institute (\citeauthor{KSRI1998}).
These two models assume an overall (resp. interframe) modal imperfection of amplitude $A_n$ (resp. $A_m$).
\par

The final semi-numerical model yielding the plastic collapse pressure of an infinite length ring-stiffened cylinder with both an overall and an interframe imperfections is approximated as follows:
\begin{equation}
	p_{\rm critical}\left(A_n,\,A_m\right) = \min\left(p_{n\,{\rm pl}}(A_n),\,p_{m\,{\rm pl}}(A_m)\right).
\end{equation}

\subsection{Formulations of the design optimization problem}

In this section two design philosophies are opposed.
The first one resorts to the so-called worst case approach that consists in designing for an extreme configuration specified by experts.
The other one uses a more comprehensive probabilistic model and eventually falls under the RBDO formulation.
\par

\subsubsection{Objective and constraints}

First, the objective of the design optimization is to find the set of parameters defining the geometry of the structure $\ve{d} = (e,\,h_{\rm w},\,e_{\rm w},\,w_{\rm f},\,e_{\rm f})\tr$ that minimizes the ratio between the structural weight and the weight of the displaced water.
The latter ratio reads as follows:
\begin{equation}
	c(\ve{d}) = \frac{\rho_{\rm steel}\,\calv_{\rm steel}(\ve{d})}{\rho_{\rm water}\,\pi\,(R + e/2)^2\,L_{\rm s}},
\end{equation}
where $\calv_{\rm steel}$ is the volume of steel composing the ring-stiffened bay and $\rho_{\rm steel} = 7,650$~kg/m$^3$ is the density of steel. 
\par

The admissible design space is bounded with the following constraints:
\begin{itemize}
	\item[\emph{(i)}] Since the semi-numerical model lacks consideration of the frame tripping collapse mode, it is proposed to resort to the following conservative safety criteria prescribed in the \citeauthor{BS5500}:
	\begin{eqnarray}
		h_{\rm w} & \leq & 1.1\,\sqrt{\frac{E}{\sigma_{\rm y}}}\,e_{\rm w}, \label{eq:frame_tripping}\\
		w_{\rm f} & \leq & \sqrt{\frac{E}{\sigma_{\rm y}}}\,e_{\rm f}.
	\end{eqnarray}
	These two constraints actually bound the slenderness ratios of the stiffener components.
	\item[\emph{(ii)}] The stiffener flange should not be too large with respect to the interframe distance:
	\begin{equation}
		445~\text{mm} \leq L_{\rm s} - w_{\rm f}.
	\end{equation}
	\item[\emph{(iii)}] The design space is bounded with the following reasonable values:
	\begin{eqnarray}
		\frac{p\,R}{\sigma_{\rm y}} \leq & e & \leq 50~\text{mm},\\
		w_{\rm f} \leq & h_{\rm w} & \leq 2\,w_{\rm f},\\
		5~\text{mm} \leq & e_{\rm w} & \leq 25~\text{mm},\\
		70~\text{mm} \leq & w_{\rm f} & \leq 150~\text{mm},\\
		15~\text{mm} \leq & e_{\rm f} & \leq 50~\text{mm}.
	\end{eqnarray}
	The first lower constraint on the hull thickness $e$ means that the circumferential stress in the equivalent non-stiffened cylinder should not exceed the yield strength.
\end{itemize}
\par

At last, the predictive models for the collapse pressure (namely the FE model and the semi-numerical solutions) are used for guaranteeing that collapse does not occur at some prescribed accidental diving depth $I_{\rm acc}$.
Therefore, it leads to the establishment of the following last constraint:
\begin{equation}
	I_{\rm acc}\,\rho_{\rm water}\,g \leq p_{\rm critical}(\ve{d}).
\end{equation}
It is assumed that the present submarine is designed for an accidental diving depth $I_{\rm acc}$ of 250~m.
\par

\subsubsection{The worst case approach}

The worst case approach basically consists in setting all the demand (resp. capacity) variables to their highest (resp. lowest) possible values and to find the optimal design for this worst scenario.
In the present context of shell design, this resorts to \emph{(i)}~prescribed maximum imperfection amplitudes and \emph{(ii)}~a destruction diving depth $I_{\rm des}$ that is significantly larger than the accidental diving depth $I_{\rm acc}$.
\par

Here, the maximum overall imperfection amplitude is taken from the \citeauthor{BS5500} recommendations and is set equal to $A_{2\,\max} = 5\,R/1,000$.
The interframe imperfection amplitude is set to $A_{14\,\max} = L_{\rm s}/100$.
The destruction diving depth is arbitrarily fixed to 340~m.
\par

\subsubsection{The probabilistic approach}

Arguing that the previous worst case approach introduces an unknown degree of conservatism, it is proposed to resort to a more comprehensive probabilistic model for describing the possible configurations of the hull.
This probabilistic model is specified in \tabref{tab:ring_stiffened_cylinder_def}.
\par

%    \centering\small
%    \renewcommand{\arraystretch}{1.15}
%    \caption{Probabilistic model for the ring-stiffened shell cylinder.}
\begin{table}
	\centering
%	\vspace{36pt}
	\caption{Probabilistic model for the ring-stiffened shell cylinder.}
	{\begin{tabular}{llcrr}
			\toprule
			\multicolumn{2}{c}{\textbf{Variable}} & {\textbf{Distribution}} & \multicolumn{1}{c}{{\textbf{Mean}}} & \multicolumn{1}{c}{{\textbf{C.o.V.}}} \\
			\hline
			$E$ & (MPa) & Lognormal & $200,000$ & $0.05$ \\
			$\sigma_{\rm y}$ & (MPa) & Lognormal & $390$ & $0.05$ \\
			$\sigma_{\rm u}$ & (MPa) & Lognormal & $570$ & $0.03$ \\
			$e$ & (mm)   & Lognormal & $\mu_e$  & $0.03$ \\
			$h_{\rm w}$ & (mm) & Lognormal & $\mu_{h_{\rm w}}$ & $0.03$ \\
			$e_{\rm w}$ & (mm) & Lognormal & $\mu_{e_{\rm w}}$  & $0.03$ \\
			$w_{\rm f}$ & (mm) & Lognormal & $\mu_{w_{\rm f}}$ & $0.03$ \\
			$e_{\rm f}$ & (mm) & Lognormal & $\mu_{e_{\rm f}}$  & $0.03$ \\[2pt]
			\hline
			$A_{2}$ & (mm)  & Lognormal & $\frac{1}{3}\,\frac{5\,R}{1,000}$ & $0.50$ \\[2pt]
			$A_{14}$ & (mm) & Lognormal & $\frac{1}{3}\,\frac{L_{\rm s}}{100}$ & $0.50$ \\[2pt]
			\bottomrule
	\end{tabular}}
	\label{tab:ring_stiffened_cylinder_def}
\end{table}

Since no data is available, the probabilistic model for the material properties is built from the recommendations available in the JCSS probabilistic modeling code \citep{Vrouwenvelder97}.
This code also prescribes a linear correlation between the yield strength $\sigma_{\rm y}$ and the ultimate stress $\sigma_{\rm u}$ in the form of a Pearson correlation coefficient $\rho = 0.75$, which is taken into account in the present analysis.
The right-skewed probabilistic model for the amplitudes of the imperfections was built with an empirical coefficient of variation of 50\% and the mean is such that the previous worst imperfections $A_{2\,\max}$ and $A_{14\,\max}$ matches the 99.5\%-quantile of the present probabilistic model.
This thus leads approximately to set the mean value equal to one third of the latter worst imperfection amplitudes as indicated in \tabref{tab:ring_stiffened_cylinder_def}.
\par

Given this probabilistic model, the original deterministic design optimization problem is transformed into a reliability-based design problem where safety is measured by means of the following failure probability:
\begin{equation}
	p_{\rm f}(\ve{d}) = \Prob{p_{\rm critical}(\ve{d},\,\ve{X}) \leq I_{\rm acc}\,\rho_{\rm water}\,g},
\end{equation}
where $\ve{X}$ is the random vector that collects all the random variables of the probabilistic model.
The optimization is performed \wrt the means of the random design variables $e$, $h_{\rm w}$, $e_{\rm w}$, $w_{\rm f}$ and $e_{\rm f}$.
The single probabilistic constraint ($n_p = 1$) reads as follows:
\begin{equation}
	p_{\rm f}(\ve{d}) \leq \Phi(-\beta_0),
\end{equation}
where $\beta_0 = 6$ in the present application (\ie $p_{{\rm f}\,0} \leq 10^{-9}$).
\par

\subsubsection{Resolution strategies}

The deterministic design optimization problem underlying the worst case approach is solved here by means of the Polak-He gradient-based optimizer.
It uses the two proposed mechanical models for the buckling strength of the structure, namely the semi-numerical (SN) and the ANM-based finite element (FE) models.
\par

The reliability-based design optimization problem underlying the probabilistic approach is solved with the metamodel-based RBDO strategy presented in \secref{sec:surrogate_based_reliability_analysis}.
Again, two designs are computed with either of the mechanical models.
\par

Once the four optimal designs are found, a reliability analysis is performed in order to compute the safety level of the optimally designed structures at both the accidental and the destruction diving depth using the probabilistic model of \tabref{tab:ring_stiffened_cylinder_def}.
Since the FE model is expensive to evaluate, this resorts to the metamodel-based importance sampling technique \citep{DubourgPEM2011} with a 5\% target coefficient of variation on the failure probability.
For the less expensive semi-numerical model, it is proposed to resort to direct subset simulation in order to compute the whole CDF of the critical pressure which yields a relationship between the failure probability and the diving depth in a single run for each design.
\par

\subsection{Results}

\begin{figure}
	\centering
	\subfigure[Worst case approach]{
		\includegraphics[width=.45\columnwidth, clip=true, trim=70 60 60 80]{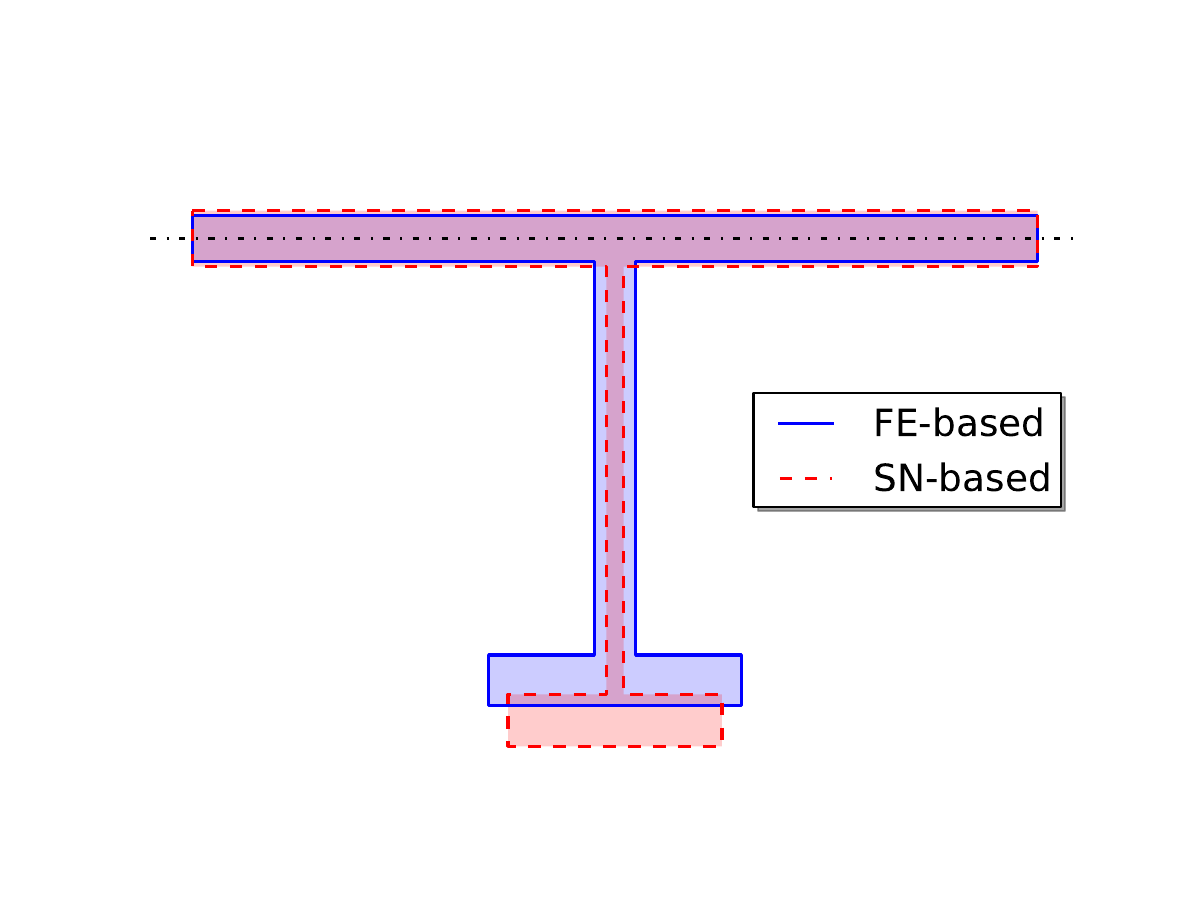}
		\label{fig:worst_case_optimal_designs}}
	\subfigure[RBDO ($\beta = 6$)]{
		\includegraphics[width=.45\columnwidth, clip=true, trim=70 60 60 80]{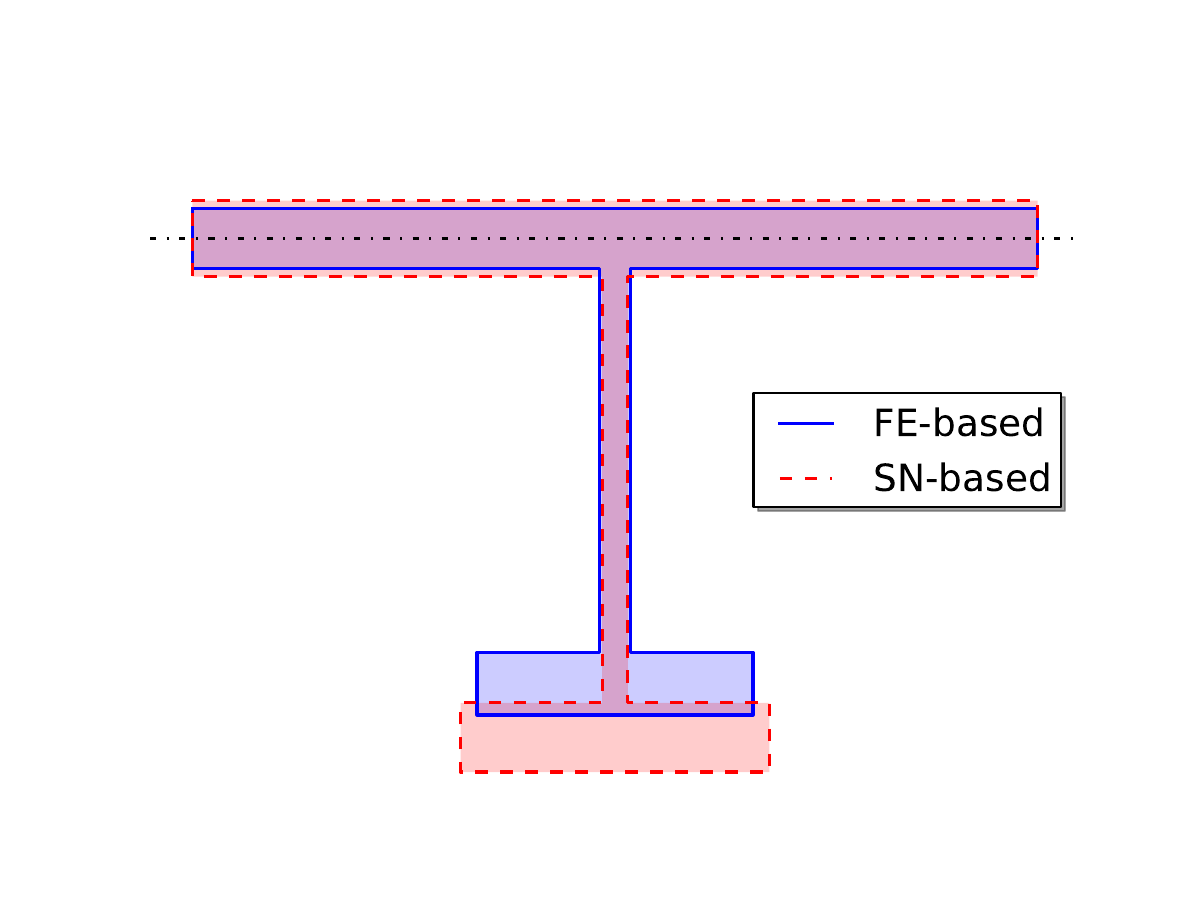}
		\label{fig:RBDO_optimal_designs}}
	\caption{Comparison of the optimal designs for the imperfect infinite-length ring-stiffened shell cylinder.}
	\label{fig:ring_stiffened_cylinder_res}
\end{figure}

The results are given in \tabref{tab:ring_stiffened_cylinder_res} and the corresponding designs are illustrated in \figref{fig:ring_stiffened_cylinder_res}.
First, it should be noticed that the FE-based design is always more cost-optimal than its SN-based counterpart.
Actually, this confirms the initial intuition as the semi-numerical solutions involve a set of built-in safety factors that eventually lead to an important (although unknown) degree of conservatism.
In the worst case approach, the relative gain in using a FE model \wrt the SN-cost is only 2\%, whereas it reaches 17\% in the RBDO approach.
\par

It should also be noticed that the SN-based design always features a more slender stiffener web than the FE-based designs.
This is because the SN-solution lacks an explicit consideration of the frame tripping buckling mode.
This lack is such that in the deterministic worst case approach, the \citeauthor{BS5500} safety constraint regarding this mode and defined in \eqref{eq:frame_tripping} is active at the optimal design.
Indeed, in this case, the stiffener web is clearly too slender as illustrated in \figref{fig:worst_case_optimal_designs}.
\par

\begin{table}
	\centering
%	\vspace{36pt}
	\caption{Results for the design optimization of the imperfect infinite-length ring-stiffened shell cylinder.}
	{\begin{tabular}{llrrrr}
			\hline
			& & \multicolumn{2}{c}{\textbf{Worst case approach}} & \multicolumn{2}{c}{\textbf{RBDO ($\beta = 6$)}} \\
			\multicolumn{2}{c}{\multirow{-2}{*}{\textbf{Method}}} & \multicolumn{1}{c}{FE-based} & \multicolumn{1}{c}{SN-based} & \multicolumn{1}{c}{FE-based} & \multicolumn{1}{c}{SN-based} \\
			\hline
			$e$ & (mm)        &  21.99 &  26.56 &  28.65 &  35.85 \\
			$h_{\rm w}$ & (mm)      & 186.01 & $^a$202.38 & 181.37 & 201.66 \\
			$e_{\rm w}$ & (mm)      &  19.47 & $^a$8.14 &  14.44 &  12.11 \\
			$w_{\rm f}$ & (mm)      & 119.57 & 101.22 & 130.62 & 146.18 \\
			$e_{\rm f}$ & (mm)      &  23.97 &  24.53 &  29.68 &  32.77 \\
			\hline
			Cost        &           &  0.1960 &  0.2004 &  0.2356 &  0.2847 \\
			\hline
			\multicolumn{2}{c}{$\beta(I_{\rm acc})$} & 4.99 & 3.81 & 6.06 & 6.11 \\
			\multicolumn{2}{c}{$\beta(I_{\rm des})$} & 1.40 & 2.00 & 4.42 & 4.99 \\
			\bottomrule
			\multicolumn{6}{l}{$^a$ The frame tripping constraint is active.}
	\end{tabular}}
	\label{tab:ring_stiffened_cylinder_res}
\end{table}

\begin{figure*}
	\centering
	\subfigure[Worst case approach]{
		\includegraphics[width=.45\textwidth,clip=true, trim=10 0 50 20]{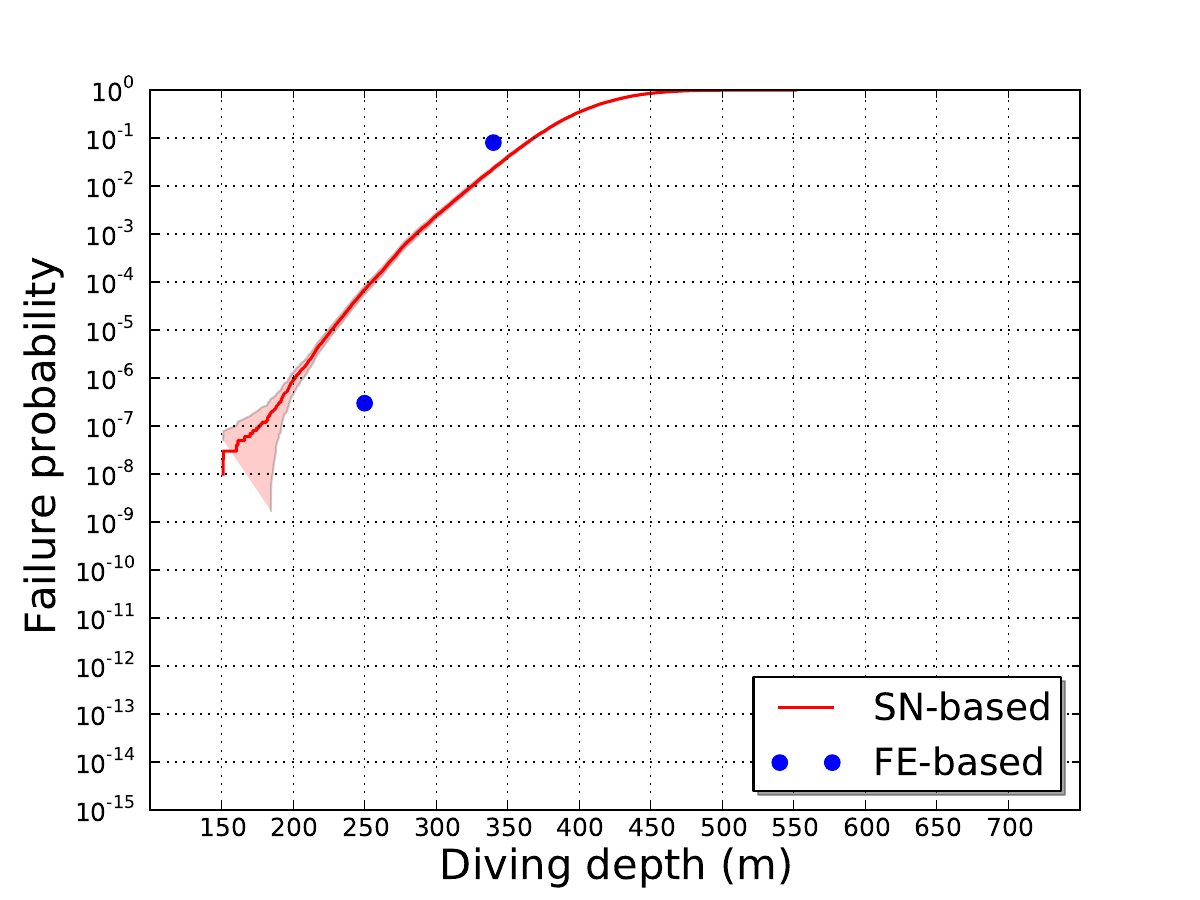}
		\label{fig:worst_case_pf_vs_depth}}
	\hfill
	\subfigure[RBDO ($\beta = 6$)]{
		\includegraphics[width=.45\textwidth,clip=true, trim=10 0 50 20]{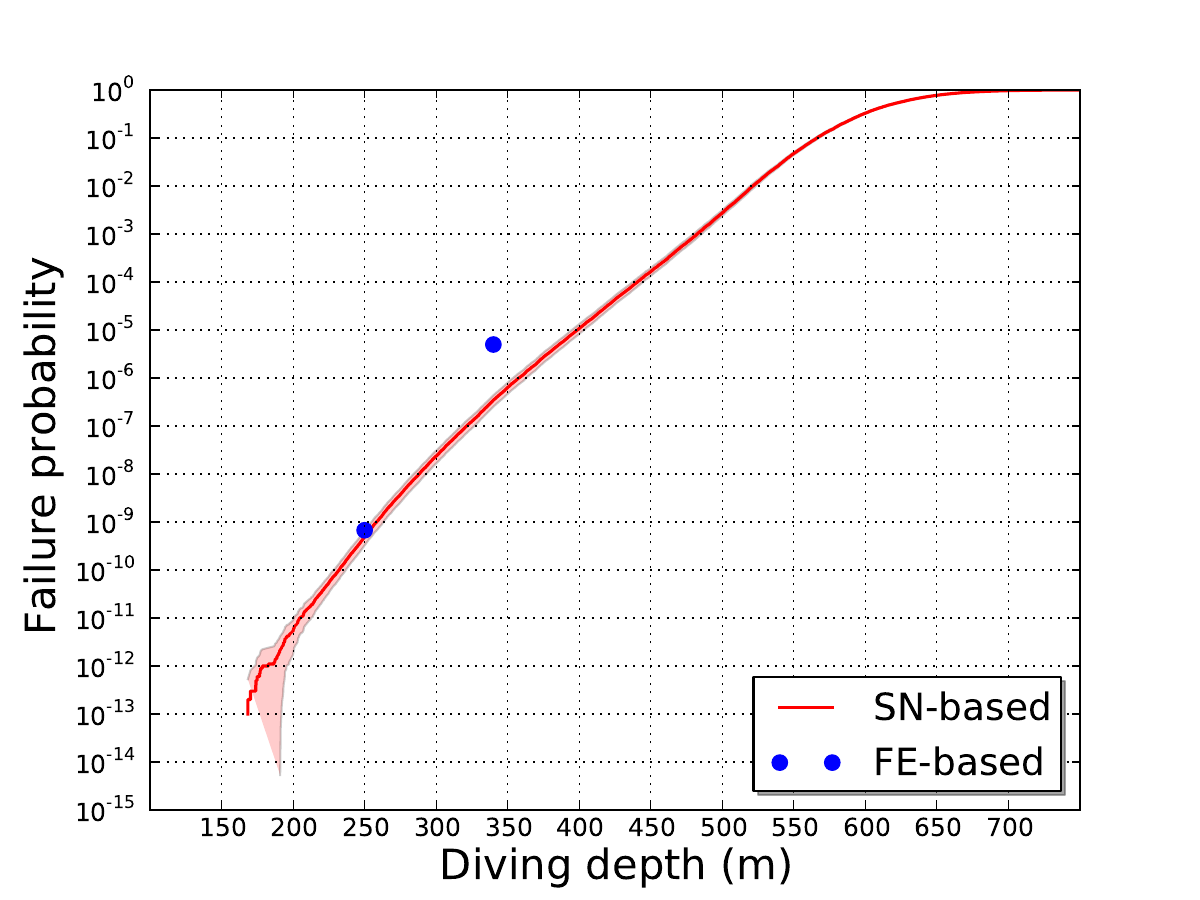}
		\label{fig:RBDO_pf_vs_depth}}
	\caption{Relation between the diving depth and the failure probability for the imperfect infinite-length ring-stiffened shell cylinder.}
	\label{fig:ring_stiffened_cylinder_res2}
\end{figure*}

As expected, the worst case approach offers a significant degree of safety at the accidental diving depth and it even remains a little margin at the destruction diving depth although the failure probability is much greater there ($p_{\rm f} \approx 10^{-2}$).
The probabilistic approach enables an explicit control of the safety level at the accidental diving depth.
Due to the important targeted level of safety ($p_{\rm f} < \Phi(-6) \approx 10^{-9}$), the reliability-based optimal designs are of course less optimal than their worst case counterparts.
\par

The relationship between the diving depth and the failure probability is illustrated in \figref{fig:ring_stiffened_cylinder_res2}.
The subset sampling technique applied with the semi-numerical model enables a reconstruction of the full CDF.
The metamodel-based importance sampling applied on the expensive-to-evaluate finite-element model only yields the failure probability estimates at the two diving depths of interest.
It can be seen from \figref{fig:RBDO_pf_vs_depth} that the failure probability matches the maximum tolerance set here equal to $p_{\rm f} = \Phi(-6) < 10^{-9}$.
\par

Convergence of the metamodel-based RBDO strategy is obtained within 850 calls to the buckling strength models.
Note that it is of utmost importance for the FE-based application due to the important numerical effort required by a single FE analysis (about 10 minutes of CPU time on our computer).
\par

%The complementary metamodel-based importance sampling analyses on the FE-based designs revealed that the Kriging surrogates accurately fit the limit-state surfaces as the correction factor is always close to unity.
%Hence, the required coefficient of variation of 5\% is obtained within a few hundred calls to the FE model.
\par

%%%%%%%%%%%%%%%%%%%%%%%%%%%%%%%%%%%%%%%%%%%%%%%%%%%%%%%%%%%%%%%%%%%%%%
\section{Conclusions}

The paper recalls the principles of an adaptive Kriging method for efficiently approximating the limit-state surfaces that appear in RBDO problems.
The Kriging variance, which is a native estimation of the surrogate precision, enables \emph{(i)}~an adaptive enrichment of the experimental design and \emph{(ii)}~the computation of approximate confidence bounds on the probabilities of failure.
The definition of the augmented space that sums the input uncertainties and the range of variation of the design parameters enables the construction and the adaptive enrichment of a unique surrogate for the whole RBDO procedure.
\par

RBDO is applied to the design of an imperfect submarine pressure hull prone to buckling. The safety margin associated with the current worst case design methodology has been quantified in the form of a failure probability.
It reveals that this common practice yields a significant level of safety although it is not truly mastered.
Second, in order to address this latter remark it is proposed to explicitly account for the uncertainties in the optimization problem.
This eventually falls under the so-called RBDO formulation which is commonly identified to be too computationally demanding for application to real industrial problems.
In this context, the proposed metamodel-based RBDO strategy truly reveals interesting to come up with a solution within less than a thousand runs of the FE model.
Note that in the case of larger target failure probabilities, the RBDO problem can be transformed so as to evaluate quantiles of the limit-state function at each step of the optimization, as recently shown in \citep{Moustapha2016SMO} in the context of car design against crashworthiness.
\par

Finally, the method could be improved so as to address geometrical uncertainties that are modeled by random fields.
This type of problems shows a larger stochastic dimension and requires algorithms more advanced that the one presented in this paper to fit the surrogates.
\par

%%%%%%%%%%%%%%%%%%%%%%%%%%%%%%%%%%%%%%%%%%%%%%%%%%%%%%%%%%%%%%%%%%%%%%
\section*{Acknowledgements}

The first author was funded by a CIFRE grant subsidized by the French national agency for research and technology (ANRT) through the company Phimeca Engineering.
\par

%%%%%%%%%%%%%%%%%%%%%%%%%%%%%%%%%%%%%%%%%%%%%%%%%%%%%%%%%%%%%%%%%%%%%%%%%%%%
%% References
%\nocite{SudretHDR}

\bibliographystyle{chicago}
\bibliography{2016_JED_DubourgBourinetSudret}
\end{document}